# Dynamic control of high-voltage actuator arrays by light-pattern projection on photoconductive switches

Vesna Bacheva[1,2], Amir Firouzeh[3], Edouard Leroy[3], Aiste Balciunaite[2], Diana Davila[2], Israel Gabay[1], Federico Paratore[2,4], Moran Bercovici[1,*], Herbert Shea[3,*], Govind Kaigala[2,5,*]

[1]Faculty of Mechanical Engineering, Technion – Israel Institute of Technology, 3200003 Haifa, Israel

[2]IBM Research Europe - Zurich, Säumerstrasse 4, 8803 Rüschlikon, Switzerland

[3]Soft Transducers Laboratory (LMTS), Ecole Polytechnique Fédérale de Lausanne (EPFL), 2000 Neuchâtel, Switzerland

[4]Current affiliation: Laboratory of Soft Materials and Interfaces, ETH Zürich, 8093 Zürich, Switzerland

[5]Current affiliation: School of Biomedical Engineering, Vancouver Prostate Centre, Life Sciences Institute, University of British Columbia, Vancouver, Canada

[*]Corresponding authors

**Abstract (250 words)**
The ability to control high-voltage actuator arrays relies, to date, on expensive microelectronic processes or on individual wiring of each actuator to a single off-chip high-voltage switch. Here we present an alternative approach that uses on-chip photoconductive switches together with a light projection system to individually address high-voltage actuators. Each actuator is connected to one or more switches that are nominally OFF unless turned ON using direct light illumination. We selected hydrogenated amorphous silicon (a-Si:H) as our photoconductive material and we provide complete characterization of its light to dark conductance, breakdown field, and spectral response. The resulting switches are very robust, and we provide full details of their fabrication processes. We demonstrate that the switches can be integrated in different architectures to support both AC and DC-driven actuators, and provide engineering guidelines for their functional design. To demonstrate the versatility of our approach, we demonstrate the use of the photoconductive switches in two distinctly different applications – control of μm-sized gate electrodes for patterning flow fields in a microfluidic chamber, and control of cm-sized electrostatic actuators for creating mechanical deformations for haptic displays.

**Introduction**
Architectures for control of low-voltage arrays are readily available and can be easily implemented in microelectronics. For example, a digital micromirror device (DMD) consists of millions of mirrors addressed with a CMOS-based control supplying 10-20 V to each pixel[1]. The low cost and availability of low-voltage CMOS makes it a viable solution not only for industrial-scale manufacturing, but also for research and prototype development. However, emerging MEMS technologies such as electrostatic[2] and field-effect actuators[3–5] rely on high-voltages, typically between hundreds and thousands of volts. Although high-voltage CMOS processes do exist, their cost remains prohibitive[6], forming a bottleneck in the development of prototyping of high-voltage actuator arrays. Current solutions for high-voltage control for such arrays rely on individual wiring of each actuator to a single off-chip high-voltage switch, such as a mechanical relay[7], a high-voltage MOSFET[8], or a photoconductive semiconductor switch[9]. Although individual wiring can be useful for small arrays, it leads to large and bulky control units that are not scalable to large arrays.

Nearly 20 years ago, Lacour *et al.*[10,11] already suggested a concept of using photoconductors to control high-voltage actuators, and demonstrated a 3×3 electroactive actuator array controlled by a laser pointer that the user manually pointed to a desired actuator. A photoconductor, typically implemented using semiconductors, is a material which when excited by light with energy higher than its bandgap, increases its charge carrier concentration leading to an increase in its conductivity[12]. After removing the light excitation, the charge carrier concentration decays, and the photoconductor becomes insulating again. To the best of our knowledge, despite of its merits, Lacour's approach was largely overlooked and only recently Hajiesmaili[13] used photoconductive nanoparticles and light emitting diodes to optically control 6×6 dielectric elastomer actuators. However, this work relies on row-column addressing, thus limiting the number of achievable patterns.

In this work, we expand the concept of on-chip photoactuation for high-voltage control and propose a more generic architecture that is based on a light projection system that runs on low-voltage electronics, which controls an otherwise passive array of photoconductive switches connected to a high-voltage supply. In this way, the low- and high-voltage circuits are entirely decoupled and communicate only through light patterns. We propose different architectures of the switches, and provide guidelines for design and scaling of such arrays. To demonstrate the versatility of our method, we implement the photoactuation for applications in two fields: reconfigurable microfluids and haptics. In the context of microfluidics, we show how the switches can be used to control an array of AC-driven (100-500 V, 25 Hz) gate electrodes embedded in a microfluidic device, allowing flow patterning[14]. For haptic applications, we demonstrate the control of an array of DC-driven (1.7 kV) electrostatic actuators[15,16], creating desired topographies. While in this work we focus on these two applications, such switching capabilities may also be useful for other high-voltage devices, such as RF devices[17] and piezo-based devices[18].

## Results and discussion
### Concept of photoactuation by light-pattern projection
Fig. 1b illustrates our concept of controlling individual high-voltage actuators in an array by using a light projection system. The system consists of two electrically separate layers: one layer consists of the actuators connected to a high-voltage power supply, with one or more photoconductors per actuator serving as switches. The second layer is an external light projection system controlled by a computer. In our work we use an LED matrix, but this could also be expanded to digital micromirror device (DMD) or LCD screens – all of which are either available commercially off-the-shelf, or can be implemented using standard low-voltage electronics. The only communication between the two layers is through the light projection, which selectively turns ON or OFF photoconductive switches and thus controls the high-voltage circuit. Fig. 1a illustrates one of the architectures used in this work. Each actuator is in parallel with one photoconductive switch (S2), and the two are in series with another photoconductive switch (S1). To turn ON the actuator, S2 remains 'closed' by not being illuminated, and S1 is illuminated resulting in an increase in its conductivity, such that most of the voltage drops across the actuator. To turn OFF the actuator, light is projected only on S2 to ensure that the actuator is grounded, and thus inactive.

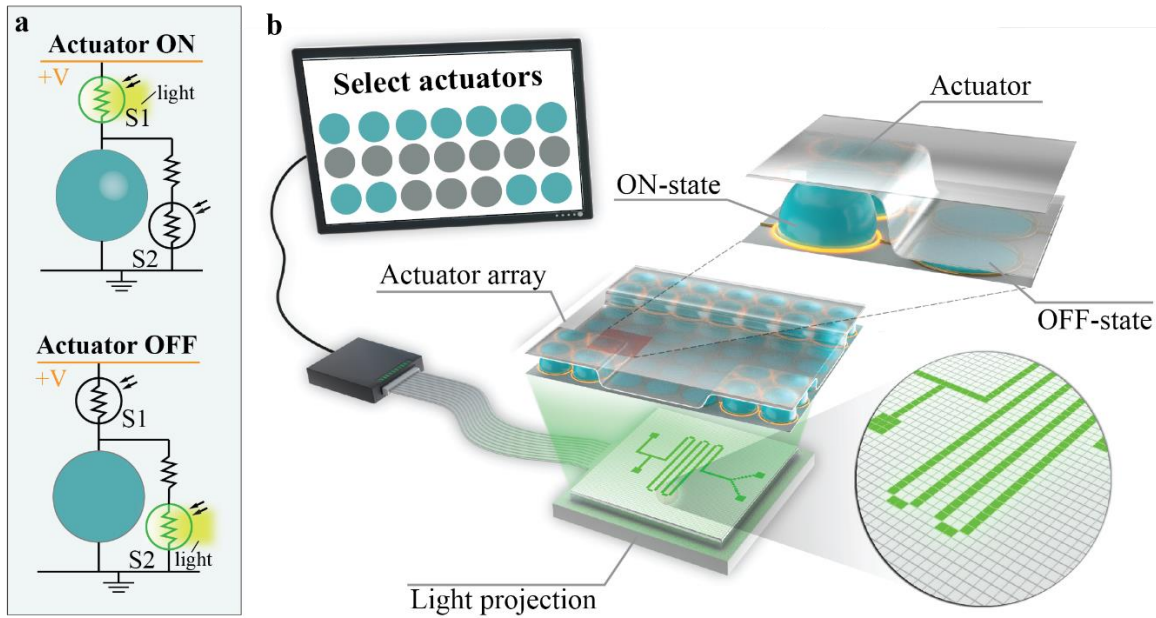

**Fig. 1 Concept of photoactuation of high-voltage actuator arrays by light projection. a** All of the actuators are connected to a single power supply through conducting voltage and ground lines. Control of each actuator in the array is enabled by two photoconductive switches S1 and S2, which increase their conductance when exposed to light. Other control architectures can be also implemented. In the presented one, to turn ON the actuator, only S1 is illuminated, such that most of the voltage drops across the actuator. To turn OFF the actuator, only $S_2$ is illuminated to ensure that the actuator is grounded. **b** Conceptual illustration of a complete photoactuation system. A light projection system (e.g., DMD, LCD screen, LED matrix) controlled by a computer is used to address the individual photoconductive switches and thus drive the actuator array.

**Design and characterization of photoconductive switches**

Central to the operation of photoconductive switches is their ability to withstand high voltages, and to maintain low dark conductance during their OFF-state (without illumination) and high light conductance during their ON-state (with illumination). Hydrogenated amorphous silicon (a-Si:H), which is widely used in solar cells and flat-panel displays, is an ideal candidate for photoconductive switches because it is known to have an electrical breakdown field of up to 16 kV/mm and a ratio of light to dark conductance of four orders of magnitude[19]. In addition, amorphous silicon can be easily deposited on various substrates (e.g. flexible[11], solid[20]), allowing integration with different types of actuators. In this work, we focus on the characterization of photoconductive switches based on a-Si:H, as it served as the material of choice for our devices. However, we conducted tests also using zinc oxide (ZnO) as a photoconductive switch, which proved to be a viable candidate for this purpose. We report the ZnO results in the Supplementary Information, Fig. S1.

Our switches are composed of metal pads (0.5 mm × 0.5 mm) spaced by a fixed gap and covered with a-Si:H as shown in Fig. 2a. To evaluate their electrical characteristics, we applied a fixed voltage across the switch and monitored the electric current in the dark and upon illumination. Fig. 2b presents a typical current response of a switch with a 100 μm gap and a 1 μm thick a-Si:H layer, showing sharp increase of the current by more than 1000-fold upon illumination with white light. While a:Si-H absorbs light in the entire visible spectrum, it is known to have the highest photoconductivity in the red region (630 - 700 nm wavelength) due to its energy band structure[21]. We observe this behavior in Fig. 2c, which presents experimentally obtained ratio of light/dark conductance as a function of the light wavelength , showing the highest ratio – more than four orders of magnitude increase – for red light. In the Supplementary Information, Fig. S1 we show similar measurements for zinc oxide (ZnO), which absorbs mostly in the UV-region and less in the visible part of the spectrum. This photoconductor could be useful for applications where visible light may need to be reserved for other uses (e.g., imaging) and UV-light could be used for triggering the switches.

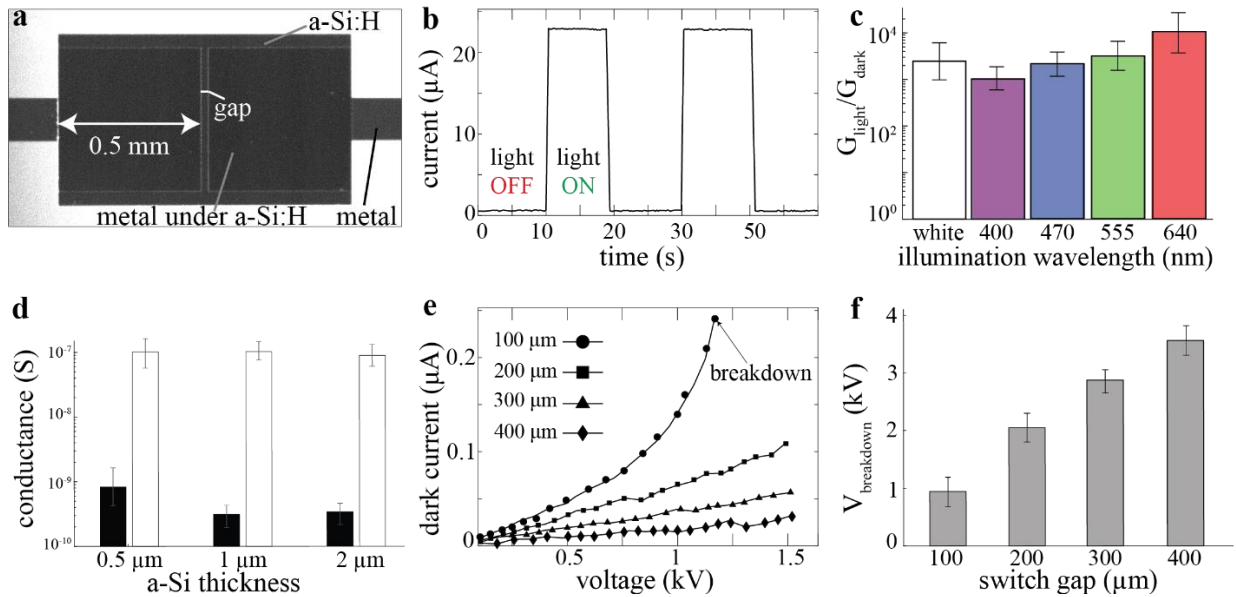

**Fig. 2 Characterization of photoconductive switches. a** Photograph of a photoconductive switch composed of amorphous silicon (a-Si:H) deposited on top of two metal pads spaced by a fixed gap. **b** Current response of a switch with a 100 μm gap length under an applied voltage of 200 V, subjected to several cycles of illumination with a white LED light (23.1 mW/cm$^2$). **c** Ratio of the light to dark conductance of a switch with a 100 μm gap length, as a function of the illumination wavelength. The ratio is normalized by the light intensity at each wavelength, showing that the red wavelength is the most effective in activating the photoconductor with a light to dark conductance ratio of 10,000. **d** Conductance of a switch with a 100 μm gap for different thicknesses of a-Si:H, in the absence (black bars) and presence (white bars) of red LED light (640 nm, 17.2 mW/cm$^2$). While the conductance at the ON-state is essentially independent of the thickness, the 0.5 μm layer exhibits much larger dark conductance and is thus less favorable. **e** Plot of the dark current as a function of the voltage for switches with different gap lengths. The current decreases with the switch gap due to the decrease in the conductance of the switch. **f** Breakdown voltage as a function of gap length, in the absence of light (OFF-state). The breakdown field in this range of gap lengths is nearly constant at approximately 10 kV/mm. The error bars in all plots represent a 95% confidence interval on the mean (with at least 10 switches, and 3 on/off cycles in each).

An ideal switch should have zero dark conductance, however due to thermally excited electrons, there is always a small current flowing even when the switch is not illuminated. Fig. 2d shows the dark and light conductance as a function of the photoconductor thickness, for a fixed 100 μm gap length. All three layers', ranging between 0.5 μm and 2 μm in thickness, exhibit similar light conductance, but the dark conductance is highest for the 0.5 μm layer most likely due to the presence of pinholes (which were visible on an optical microscope). Since the dark conductance of the 1 μm layer is slightly lower than that of the 2 μm layer, we decided to fix the thickness layer to this value for the remainder of the work. In addition, we characterized the dark current as a function of the applied voltage (up to 1.5 kV) for different gap lengths, as shown in

Fig. 2e. As expected, the current increases with the voltage and decreases with the length of the gap. For example, for 1 kV, the dark current is 50 nA for a 200 μm gap and 10 nA for a 400 μm gap.

The gap length can be tuned to accommodate for the desired application and required voltages. Clearly, the switch should be operated below its breakdown voltage, i.e., the maximal voltage that the switch can withstand in its OFF-state without short-circuiting. Fig. 2f presents the breakdown voltage for different gap lengths. As expected, the breakdown voltage increases with gap length with a nearly constant breakdown field (breakdown voltage normalized by the gap length) of approximately 10 kV/mm. These values are in agreement with reported breakdown fields of amorphous silicon[10].

**Photoactuated microfluidic device**

As a first proof of concept, we demonstrate the control of an array of AC-driven gate electrodes (100-500V, 25 Hz) that control the flow field in a microfluidic device, shown in Fig. 3a. The electrodes are deposited on the bottom of a microfluidic chamber and are separated from the liquid by a thin dielectric layer. Applying an AC voltage difference between the electrode and the liquid results in capacitive charging of the solid-liquid interface known as an electric double layer (EDL). The interaction of the EDL with an electric field parallel to the floor of the chamber gives rise to fluid motion known as electroosmotic flow (EOF)[22]. This flow patterning approach is called alternate-current field-effect electroosmosis (ac-FEEO), and is presented in detail in Paratore *et al.*[14] This approach holds the potential for creating arbitrary and dynamically configurable flow fields for microfluidics and lab-on-a-chip applications[23]. However, to date, achievable flow patterns using this method were limited by the number of electrodes that could be individually controlled.

Fig. 3 shows our implementation of a 3×3 array of individually addressable gate electrodes using photoconductive switches. Each electrode is connected to the AC power line via a single photoconductive switch that is located in the outer edges of the device. Each switch is controlled by a dedicated LED. As illustrated in Fig. 3b, the gate electrode (modeled as a capacitor composed of the EDL and the dielectric) together with the switch (modeled as variable resistor) form an RC circuit. The generated EOF velocity is proportional to the voltage drop across the capacitor. Using a 100 μm switch gap, the RC time during the OFF-state of the switch (low conductance state) is higher than the operating AC time, resulting in a low voltage drop across the capacitor and thus negligible EOF velocity. In contrast, during the ON-state, the conductivity of the switch is sufficiently high such that the RC time is much shorter than the operating AC time, allowing most of the voltage to drop across the capacitor and induce EOF. In Supplementary Information, Fig. S2 we provide experimental measurements of the RC circuit that are consistent with the observable behavior of the system.

Movie S1 presents flow patterns obtained by each of the gate electrodes sequentially, showing that all of them operate as expected and create dipole flows[14,24,25]. Movie S2 presents flow patterns obtained from several gate electrodes activated simultaneously and shows dynamic switching from one flow field to another.   Fig. 4 shows several snapshots from that video. In Supplementary Information, Fig. S3 and Movie S3 we show another architecture wherein the same gate electrode can be connected to several power supplies using multiple switches.  By illuminating the desired switch while keeping the others dark, it is possible to select the operating conditions for the gate electrodes, e.g. determine the EOF velocity.

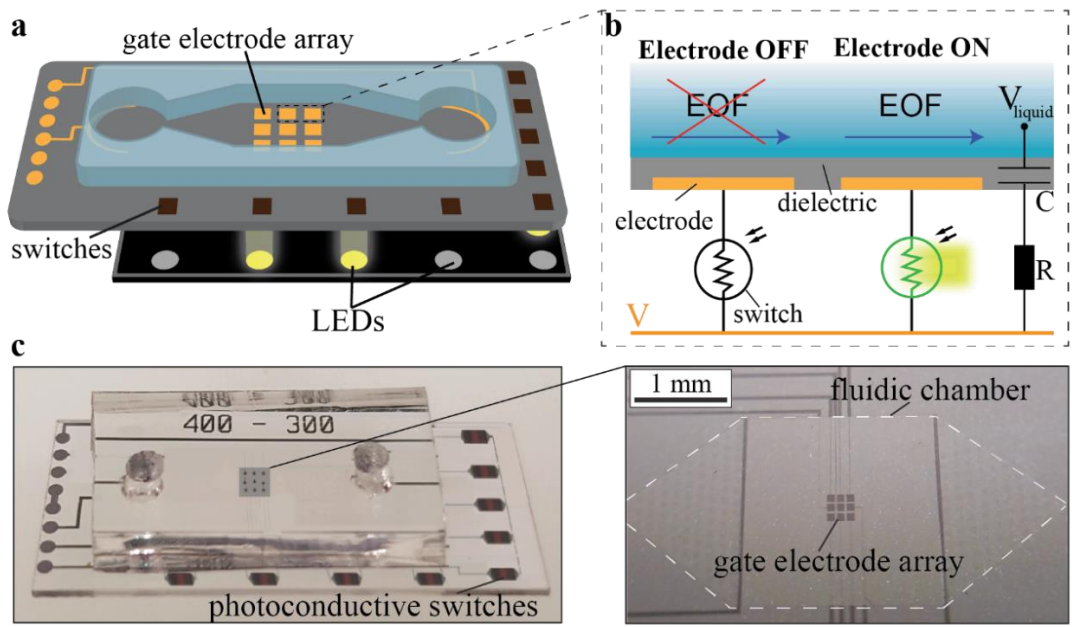

**Fig. 3 Photoactuation of gate electrodes for flow patterning. a** Illustration of a microfluidic device containing a fluidic chamber with an array of gate electrodes, and photoconductive switches with corresponding LEDs located at the edge of the device. By applying an AC potential difference between the gate electrode and the liquid, we control the electroosmotic flow (EOF) inside the chamber[14]. **b** To individually address the gate electrodes, each one is connected to the power line via a single photoconductive switch (with 100 µm gap) that can be turned ON/OFF on demand using a dedicated LED. The gate electrode and its corresponding switch form an RC circuit. In the dark, due to the high resistance of the switch, the RC time is higher than the AC time of the power supply, resulting in negligible EOF velocity. Upon illumination of the switch, the RC time decreases, allowing most of the voltage to drop across the capacitor, thus inducing EOF. **c** Photograph of the microfluidic device containing 9 photoswitches and a fluidic chamber containing an array of 3×3 gate electrodes. The inset shows a close view of the gate electrode region.

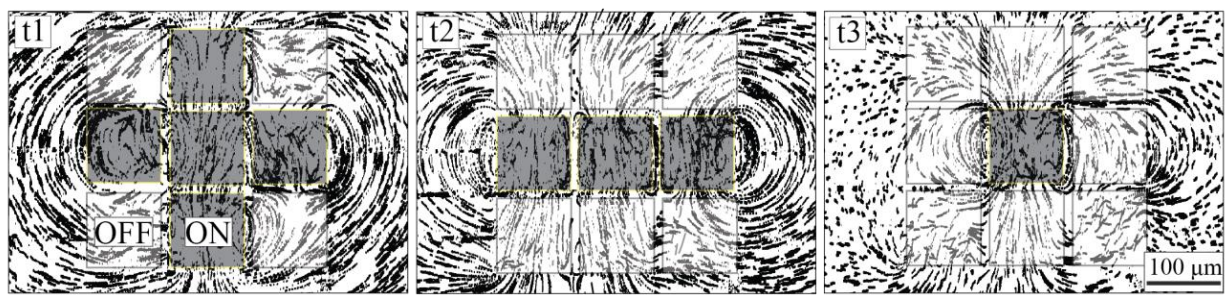

**Fig. 4 Photoactuated flow patterning.** Experimental visualization of flow streamlines generated by an array of 3×3 photoactuated gate electrodes. When an electrode is ON (gray square), its corresponding switch is illuminated, and the electrode generates local EOF. In contrast, an electrode that is OFF does not generate any flow. We switch between flow fields by illuminating different gate electrodes as shown at different times $t_1$, $t_2$ and $t_3$. A time-lapse movie showing these flow fields and the transition between them is provided in Movie S2. The electrodes are supplied with a square bipolar voltage signal (±200 V, 25 Hz) and the switches have 100 µm gap.

### Photoactuated electrostatic actuators

As a second proof of concept, we demonstrate the control of a DC-driven array of 5×5 high-voltage hydraulically amplified taxels (HAXEL), initially developed by Leroy *et al.*[15]. These electrostatic actuators are attractive for haptic application as they are capable of creating displacements of few millimeters while maintaining forces of 250 mN. They require voltages on the order of 1-2 kV. As shown in Fig. 5a, HAXEL actuators consist of two thin metallic electrodes with PET backing separated by a dielectric liquid and a dielectric layer. The top electrode has a hole at its center and is covered by an elastic membrane. When a voltage is applied between the electrodes, the electrostatic forces cause them to zip together, forcing the liquid into the central stretchable region, forming a raised bump. Removing the voltage returns the actuator to its initial position. This geometry, in which each element is independent of its neighbors, is well suited to for actuators arrays that could be useful for haptics[26], reconfigurable microfluidics[23], and adaptive optics[27], but it requires individual control of each actuator in the array.

Fig. 5b presents the implementation of the photoactuation method for controlling a HAXEL array. In contrast with the photoactuation of gate electrodes where a single switch per electrode was used, in this case each actuator is controlled by two switches $S_1$ and $S_2$. To turn OFF the actuator, only $S_2$ is illuminated, thus connecting the actuator to the ground. To turn ON the actuator, only $S_1$ is illuminated thus increasing the voltage drop across the actuator. The switch $S_1$ must be able to withstand the operating voltage of 2 kV, therefore we selected a gap of 300 μm, which has a breakdown voltage of ~ 3 kV. The light conductance of the switch $S_2$ should be significantly higher than the actuator's conductance, but its dark conductance should be significantly lower. Based on empirical test, the gap that matches this criterion is 400 μm.

Because the fabrication of the HAXEL actuators is currently not compatible with cleanroom processes used for the fabrication of the switches, we fabricated the HAXELs on separate substrates. Fig. 5c presents our method for simple integration of the two layers using a PCB that contains a set of pogo-pins on both sides – one interfacing with the actuators, and one with the actuators. The actuator and switch stack are placed around 2 mm above a computer-controlled LED matrix. In the Supplementary Information, Fig. S4-S5 we provide additional details on the design of the actuator and switch array. Fig. 6 presents an experimental demonstration of 5×5 photoactuated actuators supplied with 1.7 kV DC voltage. The illumination matrix is composed of 5×5 pairs of LEDs, where the left and right LEDs in each pair correspond to the S1 and S2 switches, respectively. Movie S4 shows the actuator array changing topography in time, in response to changes in the illumination pattern. Figure 6a presents images of the LED matrix at three points in time, each with a different actuation pattern. Figure 6b shows the actuators topography for those three cases, with the insets schematically showing the light pattern corresponding to the S1 switches.

For haptic applications, a discrete array of mechanical actuators creating an array of reconfigurable 'bumps' is ideal. However, other applications, such as reconfigurable microfluidics and adaptive optics, would benefit more from continuously deformed surfaces. Fig. 7 shows how the same array could be adapted to provide a smooth reconfigurable topography by stretching an elastic sheet on top of the HAXEL array. To visualize the deformation, we placed a ball on the membrane and demonstrate the ability to control its trajectory by light-actuating the HAXEL array. The initial state of the system is such that all actuators are ON, except for the one where the ball is positioned. As illustrated in Fig. 7a, to move the ball from one spot to another, the actuator at the origin of the ball is turned ON, while the actuator at its desired destination is turned OFF. Movie S5 shows the motion of the ball in real time, and Fig. 7b-c presents 5 time points from the video.

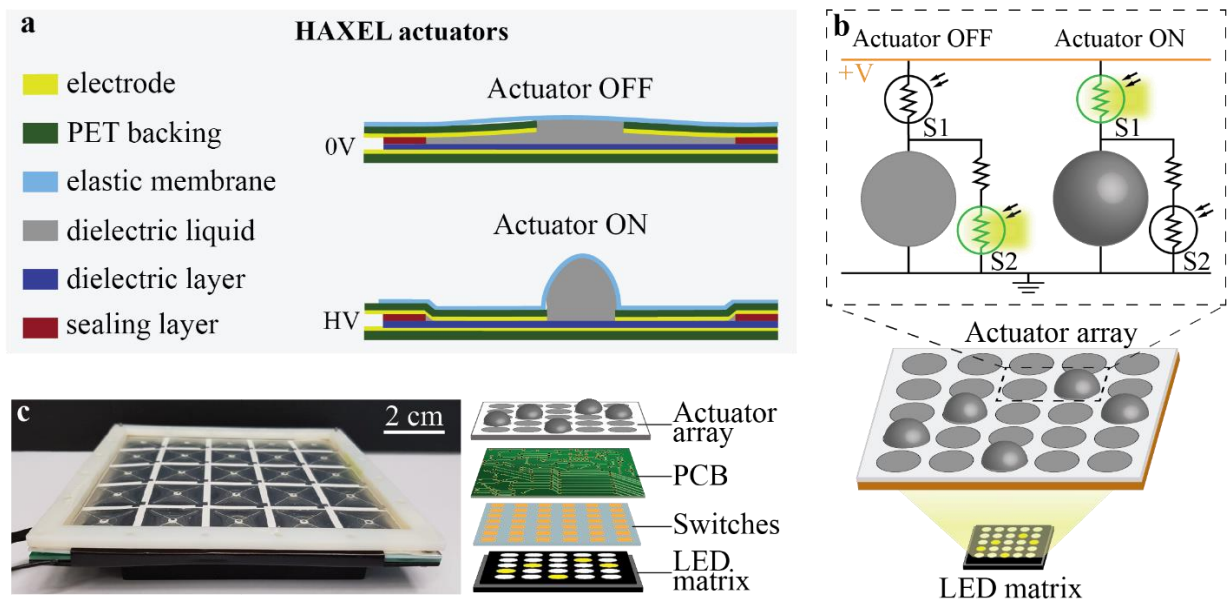

**Fig. 5 Design and architecture of a photoactuated HAXEL array. a** Structure and working principle of the HAXEL actuators. The actuators are composed of two metal electrodes with PET backing spaced by a dielectric layer and a cavity filled with a dielectric liquid. The cavity is sealed on top with elastic membrane and on the perimeter with a sealing layer. Upon application of high voltage, the electrodes zip and force the liquid into the elastic region to form a bump[15]. **b** Photoactuation architecture: each actuator is connected to two photoconductive switches $S_1$ (with 300 μm gap) and $S_2$ (with 400 μm gap). To turn the actuator OFF, the illumination to S1 is turned OFF and only S2 is illuminated, to ensure that the actuator is grounded and not charged. To turn ON the actuator, S1 is illuminated while S2 remains in the dark, such that most of the voltage drops across the actuator. **c** Photograph of the integrated device, and exploded view schematic of its four layers. A PCB board serves as the interface between the photoconductive array and the actuator array, which are all placed on top of a computer-controlled LED matrix.

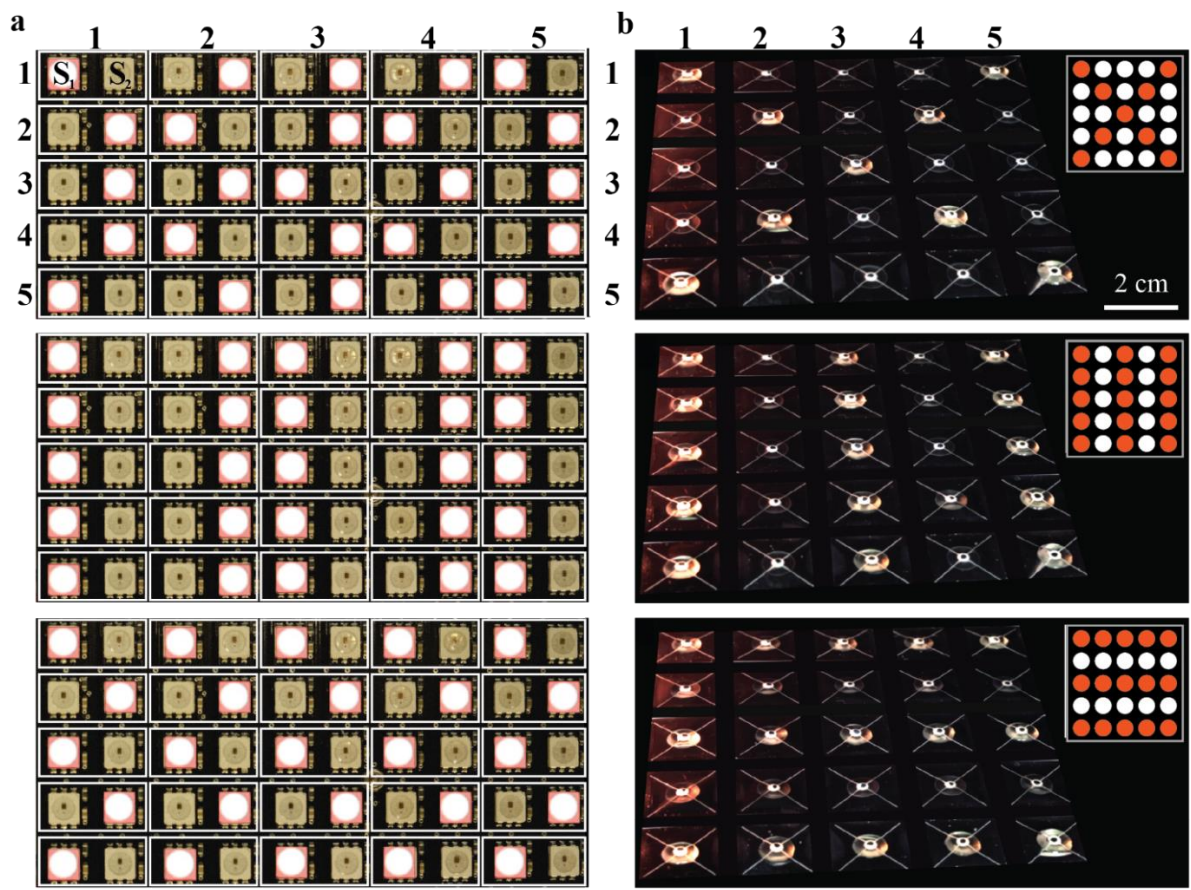

**Fig. 6 Experimental demonstration of a photoactuated HAXEL array. a** Images of the LED matrix showing three different light patterns. The matrix is composed of 5x5 pairs of LEDs, with the left one in each pair corresponding to the S1 switch and the right to the S2 switch of the actuator at location (i,j), where i and j indicate the line and row of the actuator. **b** Actuator motion resulting from each of the illumination patterns. The inset shows a schematic of the subset of LEDs corresponding to S1 switches. A time-lapse video showing these deformation patterns and the transition between them is provided in Movie S4. For better visualization, we digitally masked the regions between the actuators with a black grid.

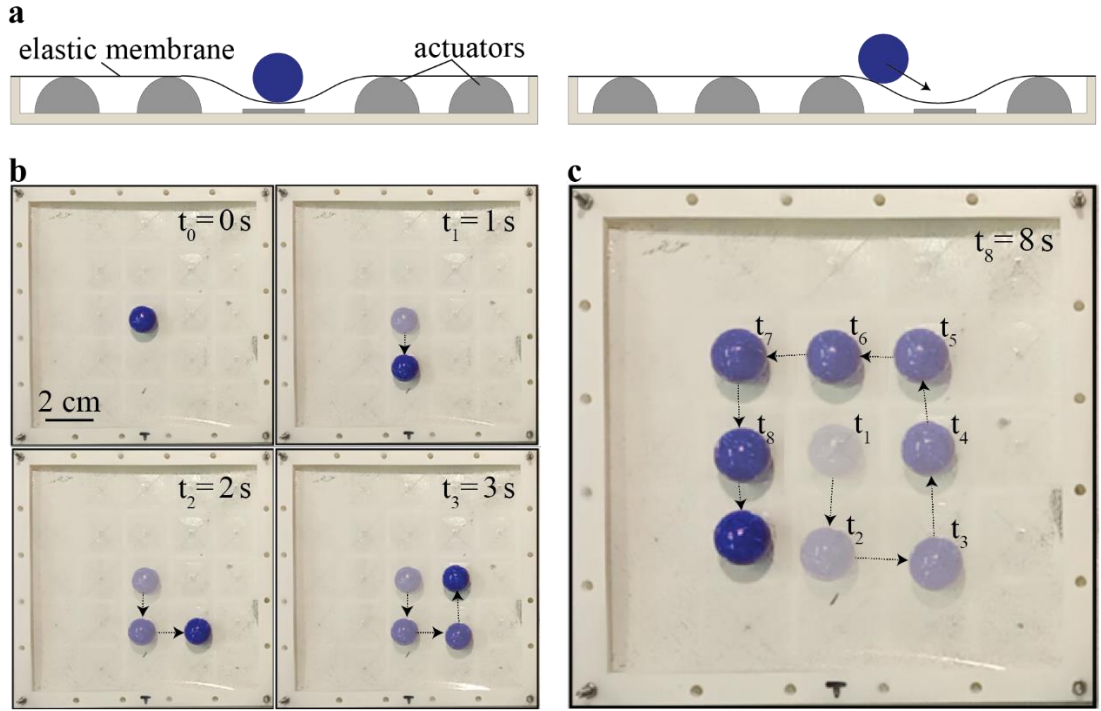

**Fig. 7 Experimental demonstration of a reconfigurable topography enabled by photoactuation. a** Schematic cross-section of a stretched membrane on top of a 5x5 photoactuated HAXEL array. We control the topography of the membrane by turning ON/OFF desired actuators, causing the membrane to deform. To visualize the deformation, we place a ball on top of the membrane and control its trajectory. To move a ball from one spot to another, we turn ON the actuator at its origin position, and turn OFF the actuator at the desired position. **b-c** Experimental demonstration of the ball moving in a trajectory dictated by the user. Each frame is superimposed with previous frames. The 5 frames are taken from a real-time Movie S5 showing the motion of the ball in real-time.

## Conclusions

We presented a method to control individual high-voltage actuators in an array using microfabricated photoconductive switches and a light projection system. In our approach, the high-voltage circuitry is electrically decoupled from the logic circuitry, allowing to use standard low-voltage electronics for controlling the array. We provided characterization of switches based on hydrogenated amorphous silicon, allowing to tailor the switch design to particular electrical requirements set by the actuator. We fabricated robust photoconductive switches with a breakdown field of approximately 10 kV/mm and a ratio of light to dark conductance of more than 10,000, and showed their utility in two very different applications – controlling AC-driven gate electrodes for generating EOF and controlling DC-driven HAXEL actuators for creating spatial topographies. We also showed that the photoconductive switches could be arranged in different architectures to support these applications. e.g. a single-switch architecture for AC-based actuation and a two-switch architecture for DC-based actuation.

In this work we used relatively small array sizes as a proof of concept, however we see no fundamental limitation in scaling-up the approach to much larger arrays. The microfabrication of the photoconductive switches is very robust and once the deposition process was optimized, we consistently observed a very high fabrication yield. With the current design, the footprint of the switch is approximately 1 mm$^2$, which can readily serve for applications such as Braille display where the required dot diameter is 1.5 mm and the distance between dots is around 2.4 mm.[28] Such arrays could be controlled either by an LED matrix as we have done in this work, or with off-the-shelf illumination units such as projectors. Flat-panel screens might also work for this application and be even more compact, provided that they provide sufficient light intensity for activating the photoconductors. For microfluidic applications, further reduction in the size of the photoconductive switches would be required. The aspects to consider and optimize are the area required for good ohmic contact of the metal pads with the photoconductive material, and the minimal gap length that can be sustained without breakdown. Much optimization can be done with amorphous silicon, but it is likely that other photoconductive materials such as silicon carbide[29] and gallium nitride[30] could provide superior performance in terms of breakdown resistance and thus allow further miniaturization.

**Materials and methods**
**Fabrication of the photoconductive switches**
To create the photoconductive switches, we first fabricated the metal layer consisting of conducting lines with gaps where a photoconductive switch is to be created and of contact pads. The metal layer consisted of 5 nm Ti/ 30 nm Ni/ 5 nm Ti. It was deposited on a 0.5 mm-thick, 4' wafer double-polished borosilicate wafer (Plan Optik AG, Germany) by photolithographic patterning followed by e-beam evaporation (BAK501 LL, Evatec, Switzerland) and a lift-off process. We then deposited amorphous silicon on the entire wafer. To that end, we optimized a plasma-enhanced chemical vapor deposition process, resulting in a highly uniform layer with < 4 % non-uniformity as measured by an ellipsometer (FilmTek SE, Bruker, USA) and a deposition rate of 39.06 nm/min. We performed the deposition using a 100 PECVD System (Oxford PlasmaPro, England) with the following process conditions: power of 50 W, temperature of 350 °C, pressure of 1800 mTorr, mixture of $SiH_4$/He (2%/98%) at a flow rate of 500 sccm and Ar at a flow rate of 400 sccm. Finally, to define the footprint of the switches, we removed the amorphous silicon everywhere except for well-defined regions around the gaps in the conducting lines. We used reactive ion etching (Oxford RIE, England) using the following process conditions: power of 60 W, pressure of 50 mTorr, $SF_6$ flow rate of 100 sccm and Ar flow rate of 100 sccm, resulting in etch rate of 166 nm/min.

**Fabrication of the microfluidic device**
For the microfluidic device, we first fabricated the metal layer consisting of gate electrodes, conducting line of the switches and pads used to interface the device with power supplies. For that, we deposited 5 nm Ti/ 30 nm Pt/ 5 nm Ti on a 0.5 mm-thick, 4' wafer double-polished borosilicate wafer (Plan Optik AG, Germany) by photolithographic patterning followed by e-beam evaporation (BAK501 LL, Evatec, Switzerland) and a lift-off process. We then deposited and structured the amorphous silicon as described in the previous section. After that, we deposited over the entire wafer a dielectric layer composed of 500 nm SiON and 100 nm of $SiO_2$ by plasma-enhanced chemical vapor deposition (100 PECVD System, Oxford PlasmaPro, England). Using photolithographic patterning, we created exposed regions of the dielectric at specific locations for electric connection (driving, ground, and pads), and we used buffered hydrofluoric acid (BHF) etching to remove the dielectric at those locations. Finally, we constructed the microfluidic chamber walls using a 15 µm thick layer of SU8, and we sealed the device with 2 mm thin PDMS slab containing openings for the reservoirs.

**Fabrication of the actuators**

For the HAXEL array, we first defined the layout of top and bottom electrodes on aluminized PET (30 nm of Aluminum on 12 µm and 50 µm PET for the top and bottom layer, respectively) by photolithographic patterning. We then cut the aluminized PET using laser micro machining (Trotec Speedy 300, Austria). On top of the bottom electrodes, we applied a solid dielectric layer by blade-casting (Zehntner ZAA 2300 film applicator, Switzerland) with Methyl Ethyl Ketone (MEK) as solvent. The dielectric layer is 35 µm thick and is composed of 70% Barium Titanate ($BaTiO_3$) particles (~3 µm) and 30% PVDF-HFP (by weight). On top of the top electrode, we bonded a silicone membrane using a combination of plasma activation (Diener ATTO – ZEPTO Plasma System, Germany) and silanization with (3-Aminopropyl)triethoxysilane (APTES). We bonded together the top and bottom electrodes using an adhesive layer, and used a laser cutter to create opening of the cavities formed between the two layers. Finally, we filled each cavity with FR3 dielectric oil (vegetable oil).

**Experimental setups**

<u>Characterization of the switches</u>. We used a four-color (violet, cyan, green and red) LED (Mira, Lumencor, USA) as a light source, and measured the light intensity for each color using an optical power meter (PMKIT, Newport, USA). The use of 'white light' refers to light obtained by turning on the four colors simultaneously. We used a high-voltage power supply (Keithley 2410, Tektronix, USA) and a custom MATLAB code (R2019b, Mathworks) to apply the voltage and to record the dark and light currents.

<u>Photoactuated microfluidic device</u>. We used 1 µm-diameter pink carboxyl polystyrene particles (Spherotech Inc., USA), mixed in a buffer composed of 10 mM acetic acid and 1 mM NaOH (Sigma-Aldrich, Switzerland), as tracer particles to visualize the flow. For visualization of the particles, we used an upright fluorescence microscope (AZ100, Nikon, Germany) equipped with a solid-state light source (Mira, Lumencor, USA), a 5× objective (AZ-Plan Fluor, Nikon, Germany) and an mCherry filter cube (562/40 excitation, 641/75 nm emission, and 593 nm dichroic mirror, Nikon, Germany). We imaged using a CCD camera (Clara, Andor-Oxford Instrument, UK) with and exposure time of 100 ms. The gate electrodes and the driving electrode were actuated with two power supplies (Keithley 2410, Tektronix, USA) producing square-wave signals (-200, 200 V) at a frequency of 25 Hz. We used an in-house MATLAB code (R2019b, Mathworks) that sets the alternating voltages of the power supplies (-200, 200 V), and a pulse generator (Stanford Research Systems, DG535) that sends a square-wave signal to the power supplies and triggers the switching of the alternating voltages. The photoconductive switches had 100 µm gap and were illuminated with dedicated LEDs (RND 135-00129, RND Electronics, China) controlled by manual switches (RND 210-00189, RND Electronics, China).

<u>Photoactuated HAXEL actuators</u>. We used two power supplies (Keithley 2410, Tektronix, USA) connected in series to provide a DC voltage of 1.7 kV. The photoconductive switch $S_1$ had a 300 µm gap, and $S_2$ had a 400 µm gap. We used a red LED matrix (80-LED RGB Matrix VM207, Velleman, Belgium) controlled by an Arduino Uno (SMD R3, Arduino, Italy) and a custom Python code running on a computer as a light projection system.

**Acknowledgements**

We thank S. Shmulevich for useful discussion and continuous support; U. Drechsler for continuous help and support on the microfabrication; the Micro and Nano Fabrication and Printing Unit at Technion where zinc oxide deposition was performed, and particularly G. Ankonina for his help in developing the deposition process; H. Wolf for recording the actuator deformations in Movie S4 with his photography equipment; V.B., and G.V.K. acknowledge R. Allenspach, H. Riel, and W. Reiss for their continuous support. We gratefully acknowledge funding from the Israel Science Foundation grant no. 2263/20 and the Swiss National Science Foundation grant no. 200021_200641.

Page **13** of **15**

29. Kelkar, K. S., Islam, N. E., Fessler, C. M. & Nunnally, W. C. Design and characterization of silicon carbide photoconductive switches for high field applications. *J. Appl. Phys.* **100**, 124905 (2006).
30. Leach, J. H., Metzger, R., Preble, E. A. & Evans, K. R. High voltage bulk GaN-based photoconductive switches for pulsed power applications. in *Gallium Nitride Materials and Devices VIII* vol. 8625 294–300 (SPIE, 2013).
Page **15** of **15**

# SUPPORTING INFORMATION

## Dynamic control of high-voltage actuator arrays by light-pattern projection on photoconductive switches

Vesna Bacheva[1,2], Amir Firouzeh[3], Edouard Leroy[3], Aiste Balciunaite[2], Diana Davila[2], Israel Gabay[1], Federico Paratore[2,4], Moran Bercovici[1,*], Herbert Shea[3,*], Govind Kaigala[2,5,*]

[1]Faculty of Mechanical Engineering, Technion – Israel Institute of Technology, 3200003 Haifa, Israel

[2]IBM Research Europe - Zurich, Säumerstrasse 4, 8803 Rüschlikon, Switzerland

[3]Soft Transducers Laboratory (LMTS), Ecole Polytechnique Fédérale de Lausanne (EPFL), 2000 Neuchâtel, Switzerland

[4]Current affiliation: Laboratory of Soft Materials and Interfaces, ETH Zürich, 8093 Zürich, Switzerland

[5]Current affiliation: University of British Columbia, Vancouver, Canada

[*]Corresponding authors

**Contents**

S1. Characterization of ZnO

S2. Time response of photoactuated FEEO gate electrodes

S3. Photoactuated HAXEL array

S4. Captions for movies

S5. References

## S1. Characterization of ZnO

In this work, we used hydrogenated amorphous silicon (a-Si:H) as a photoconductor material. However, we conducted some of our initial tests also using zinc oxide (ZnO). Specifically, we measured the ratio of light to dark conductance as a function of the illumination wavelength and compared it with that of a-Si:H, as shown in Fig. S1. While a-Si:H responds to light in the entire visible spectrum, ZnO responds significantly stronger to UV-light (e.g. a 1000-fold more than to red light) due to its energy band structure[1]. This may be advantageous in applications where decoupling between actuation and imaging is desired (e.g., actuation in UV with fluorescence imaging in the red).

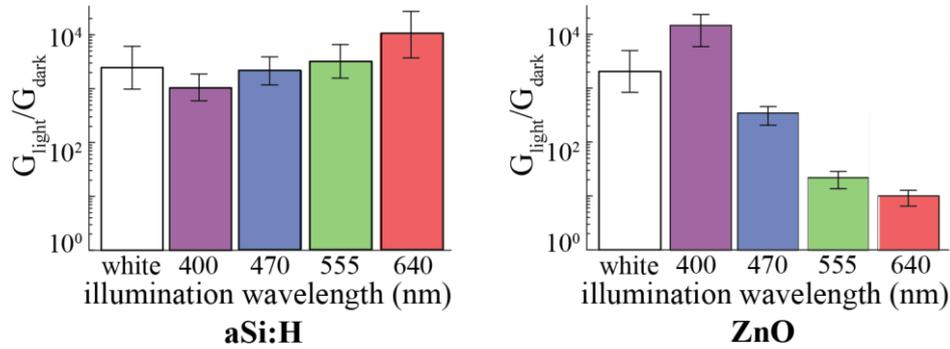

**Fig. S1.** Comparison of the the light to dark conductance ratio as a function of illumination wavelength for a-Si:H and ZnO photoconductive switches. The power density of the light was measured using optical power meter and is: 73.1 mW/cm$^2$ for white, 17.2 mW/cm$^2$ for red, 23.3 mW/cm$^2$ for green, 18.1 mW/cm$^2$ for cyan, 16.5 mW/cm$^2$ for violet, and the presented conductance ratio is normalized by the light intensity at each wavelength. Both switches had a 100 µm gap and were subjected to 200 V. We subjected each switch to three cycles of 10s in the dark, followed by 10s upon illumination, and based on the obtained results, we calculated an average dark and light conductance. The error bars represent the 95% confidence interval of the mean (with at least 10 switches, and 3 on/off cycles in each).

## S2. Time response of photoactuated FEEO gate electrodes

As described in the paper, the gate electrode together with the switch can be regarded as an RC circuit for their electrical response. The switch should operate such that in the absence of illumination, due to the low conductance in the switch, the resulting RC time is higher than the operating AC time, resulting in a low voltage drop across the capacitor and thus negligible EOF velocity. In contrast, during the ON-state, the conductivity of the switch is sufficiently high such that the RC time is much shorter than the operating AC time, allowing most of the voltage to drop across the capacitor and induce EOF.

We here estimate the RC time of a gate electrode connected to a photoconductive switch. Fig. S2 shows a schematic of the system that can be modeled as an RC circuit[2]. The capacitor C is composed of the capacitance of the electric double layer ($C_{EDL}$) in series with the dielectric capacitance $C_d$. The resistor R is composed of the resistance of the liquid channel ($R_{channel}$) in series with the resistance of the switch ($R_{switch}$), which decreases when exposed to light. We calculate the values of $C_{EDL}$ using the relation[2]

$$C_{EDL} = \frac{\varepsilon_l A}{\lambda_{EDL}},$$

where $\varepsilon_l$ is the dielectric constant of the liquid, A is the area of the gate electrode (200 µm in diameter), and $\lambda_{EDL}$ is the thickness of the electric double layer. For our buffer (10 mM acetic acid and 1 mM NaOH), the $\lambda_{EDL}$ is approximately 10 nm.[3] Similarly, we calculate the dielectric capacitance $C_d$

$$C_d = \frac{\varepsilon_d A}{d},$$

where $\varepsilon_d$ is the dielectric constant of the thin SiON layer, and d (600 nm) is its thickness. Using these relations, we estimate that the total capacitance of the system is approximately 10 pF. We measured the resistances in the system and obtained values of 150 MΩ, 15 GΩ, and 10 MΩ for the channel's resistance, the switch's dark resistance, and the switch's light resistance, respectively. Using these values for C and R, we obtain that the resulting RC time of our system in the dark is approximately 150 ms, and upon illumination is approximately 1.6 ms.

The voltage drop over the capacitor subjected to an AC voltage $V_s(t) = V_s\cos(\omega t)$ is given by[4]:

$$V_c(t) = V_s \frac{1}{\sqrt{1 + (RC\omega)^2}} \sin(\omega t + \varphi),$$

where $\omega$ is the operating frequency, and $\varphi$ is a phase constant ($\tan(\varphi) = 1/RC\omega$). The time-average voltage drop over the capacitor is given by:

$$V_{c,avg} = \omega \int_0^{1/\omega} V_c(t)dt.$$

Considering that we use an AC voltage with a 25 Hz operating frequency (i.e., 40 ms period), the average voltage drop over the capacitor is only 12 % of $V_s$ in the absence of illumination, and 92 % of $V_s$ when the switch is illuminated. This allows the switch to be effective in AC, and is consistent with the observations in our experiments.

A single gate electrode can be also connected to several switches as shown in Fig. S3. Each switch is connected to a different power supply. By illuminating the desired switch while keeping the others dark, it is possible to select the operating voltage for the gate electrodes, and thus the EOF velocity.

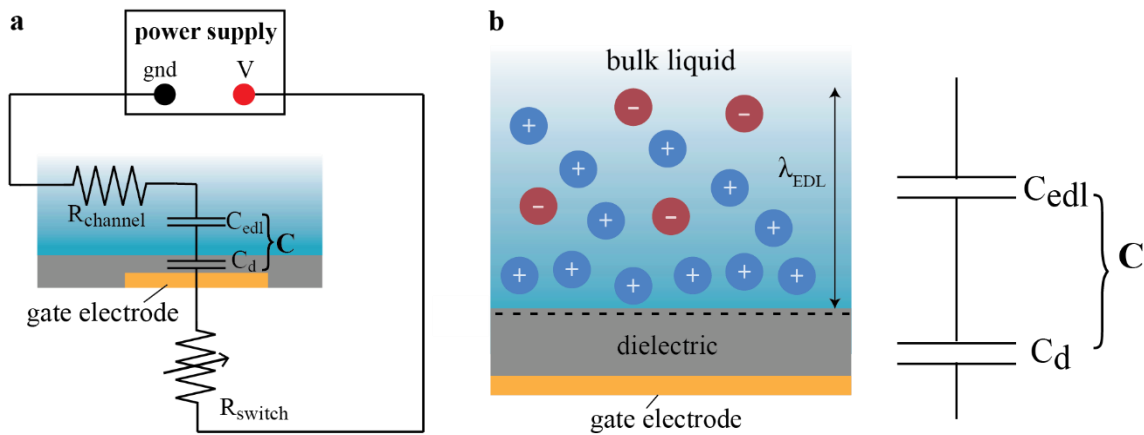

**Fig. S2.** RC model of the gate electrode in series with a photoconductive switch. The capacitor C is composed of the capacitance of the electric double layer ($C_{EDL}$) in series with the dialectic capacitance $C_d$. The resistor R is composed of the resistance of the liquid channel ($R_{channel}$) in series with the resistance of the switch ($R_{switch}$). **b** Schematic of the electric double layer on top of a gate electrode, and their equivalent capacitance.

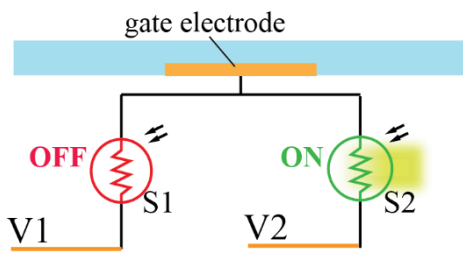

**Fig. S3.** Schematic illustration of a single gate electrode connected to two power supply via two switches S1 and S2. By illuminating the desired switch, the gate electrode is supplied with the corresponding voltage. We show two power supplies for simplicity, but this concept is extendable to multiple supplies.

## S3. Photoactuated HAXEL array

Fig. S4a shows a 3D rendering of the different layers of a 5x5 HAXEL array. Fig. S4b shows a photograph of a fabricated HAXEL array, and the LED matrix used for its control. Fig. S5 shows the array of photoconductive switches that are located under the array.

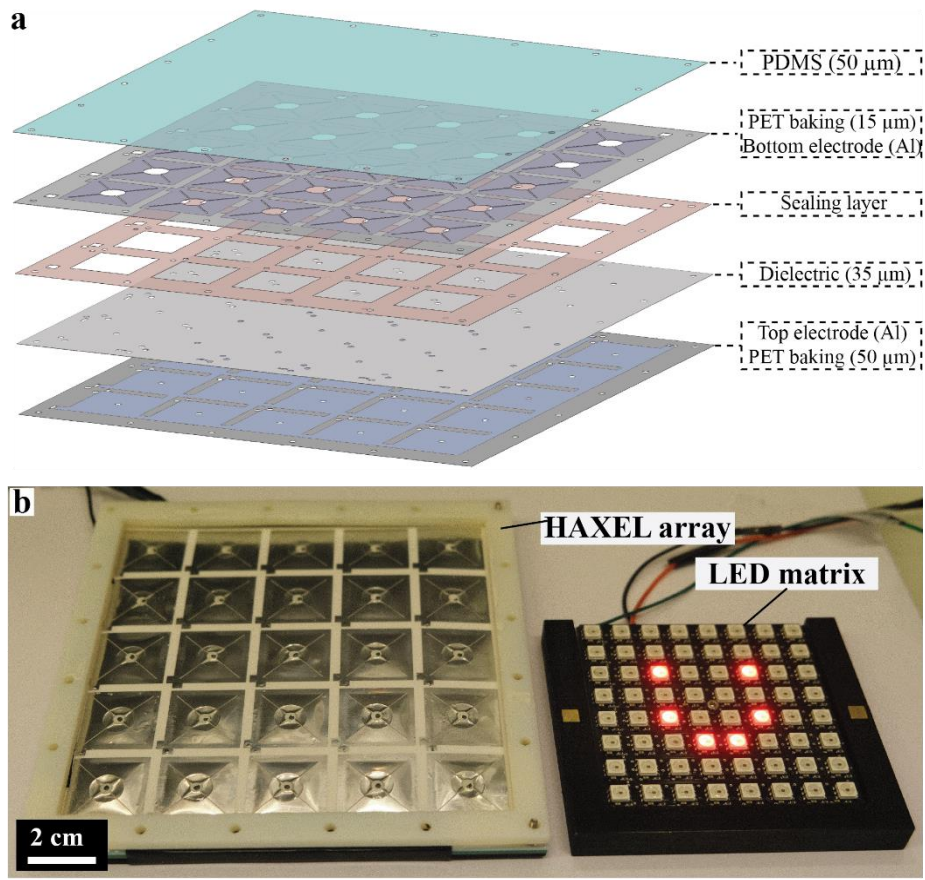

**Fig. S4 a** Exploded view of a 5x5 HAXEL array that is composed of five main layers. **b** Photograph of the HAXEL array and the LED matrix, which normally it is fixed under the array using magnets.

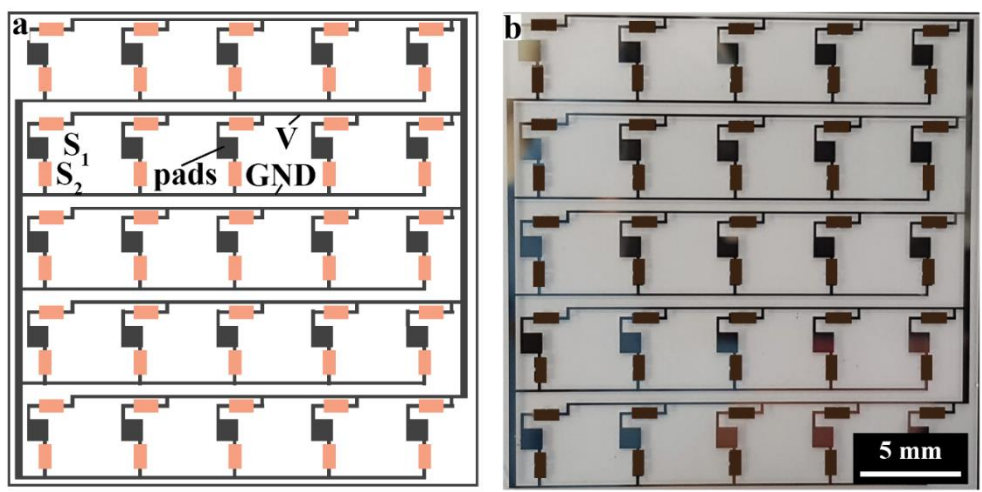

**Fig. S5. a** Schematics of the layout of the photoconductive switch used for the HAXEL array. Each actuator is connected to two switches S1 and S2. All $S_1$ switches are connected to the power line (voltage V), and all switches $S_2$ are connected to a common ground. Metallic pads are used to interface with the actuator array. **b** Photograph of microfabricated photoconductive switch array.

## S4. Captions for movies

**Movie S1. Individual addressing of gate electrodes.** A single square-shaped gate electrode gives rise to an electroosmotic dipole flow[2,5]. The yellow square indicates which electrode is turned ON by illumining its corresponding switch. Each frame of the video was background subtracted and obtained by the superposition of 5 frames of the raw data time-lapse.

**Movie S2. Photoactuated flow patterns with one power supply.** Flow patterns obtained by actuating several gate electrodes simultaneously. Each frame of the video was background subtracted and obtained by the superposition of 5 frames of the raw data time-lapse.

**Movie S3. Photoactuated flow patterns with two power supplies.** Flow patterns generated by several photoactuated gate electrodes connected to two power supplies generating flow upward (red square) or downward (blue square). By using two switches per electrode, we can select which electrode is ON and the direction of the flow on each electrode. Each frame of the video was background subtracted and obtained by the superposition of 5 frames of the raw data time-lapse.

**Movie S4. Photoactuated HAXEL array.** Deformations obtained by a photoactuated 5x5 HAXEL array. The inset shows the light patterns of S1 switches only, and indicates which actuator is ON. For better visualization, we digitally masked the regions between the actuators with a black grid.

**Movie S5. Motion of a ball on a programmable topography.** The trajectory of the ball is dictated by the topography of an elastic membrane, which is suspended on top of a 5x5 HAXELs array. To move the ball form one point to another, we turn ON the actuator at its origin, and turn OFF the actuator at its desired location. The inset shows the light patterns corresponding to S1 switches only.